\documentclass[letterpaper]{article}
\usepackage{aaai}
\usepackage{times}
\usepackage{helvet}
\usepackage{courier}
\usepackage{graphicx}
\usepackage{soul}
\usepackage[colorinlistoftodos]{todonotes}
\usepackage{eurosym}
\begin{document}
%
\title{Ranking with Social Cues: \\  Integrating Online Review Scores and Popularity Information}
\author{Pantelis P. Analytis\footnote[1], Alexia Delfino\footnote[2], Juliane E. K\"{a}mmer\footnote[3], Mehdi Moussa\"{i}d\footnote[3], Thorsten Joachims\footnote[1]\\
\\
\\ Cornell University\footnote[1]{} \hspace{3mm} London School of Economics\footnote[2]{} \hspace{3mm} Max Planck Institute for Human Development\footnote[3]{} \\
}
\maketitle
\begin{abstract}
\begin{quote}

Online marketplaces, search engines, and databases employ aggregated social information to rank their content for users. Two ranking heuristics commonly implemented to order the available options are the average review score and item popularity---that is, the number of users who have experienced an item. These rules, although easy to implement, only partly reflect actual user preferences, as people may assign values to both average scores and popularity and trade off between the two. How do people integrate these two pieces of social information when making choices? We present two experiments in which we asked participants to choose 200 times among options drawn directly from two widely used online venues: Amazon and IMDb. The only information presented to participants was the average score and the number of reviews, which served as a  proxy for popularity. We found that most people are willing to settle for items with somewhat lower average scores if they are more popular. Yet, our study uncovered substantial diversity of preferences among participants, which indicates a sizable potential for personalizing ranking schemes that rely on social information. 
%
%

\end{quote}
\end{abstract}


Online venues regularly use social information to rank or organize their content for users. Best-seller, most-read, and most-cited lists are widely implemented in online journals, newspapers, and  scholarly search engines to rank contents. Similarly, online marketplaces such as Amazon offer users the option to rank products by their average score or the number of reviews. Social information can improve user choices  on these websites, as it summarizes various attributes to a small, yet accurate, feature vector. At the algorithmic level, it facilitates preference learning and ranking for the same reasons: algorithms can learn user preferences for small feature vectors much faster.

In this paper, we study how people trade off between two widely used information sources---the average review score and the popularity of an item. Intuitively, we would expect people to integrate information about these two features. Little is known, however, about actual user preferences. How do users integrate these two pieces of information and what is the best functional form to capture their preferences? Answers to these questions would inform the design of composite social ranking mechanisms that leverage the information contained in both social cues. 

There are several good arguments for using item popularity, in addition to average review scores, to infer product quality. First, as people can give only noisy estimates about an objective true value, aggregating the opinions of several individuals reduces the estimation error \cite{galton1907vox}. Moreover, people have diverse tastes. If many other individuals have enjoyed an option, it may be more likely that a user will also do so. A low number of reviews, in contrast, could indicate a niche product that is enjoyed only by certain user communities. Further, because options with higher average review scores stand a better chance of being selected, product owners may benefit from faking reviews. It has been well documented that review manipulation takes place on many online venues, such as Yelp \cite{luca2016fake}. Yet, the extent of manipulation seems to be moderate. It is thus reasonable to expect that options with a large number of reviews are more robust against it---their average review score will be less diluted by fraudulent reviews.

The previous arguments were inferential in nature, assuming that popularity signals quality. Yet, people might have an inherent preference for more popular options over less popular ones. Some people may be eager to share their experiences with others and, for instance, enjoy discussing cultural products with friends or colleagues \cite{watts2011everything}. Others may prefer to conform with other people's choices as a way to fit into a social group \cite{asch1951effects}. Finally, less popular products may be more risky, as their actual utility remains more uncertain. Users who are risk averse and want to avoid negative experiences might discount the value of options with a low number of reviews.

\begin{figure}[htbp]
\includegraphics[width=\columnwidth, height=0.27\textheight]{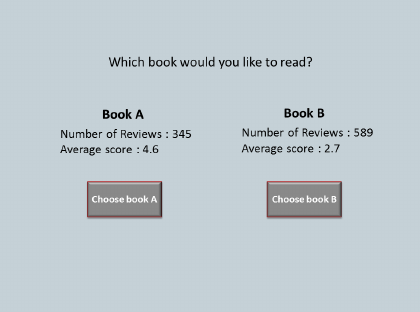}
\caption{Experimental design: The two options appeared next to each other. Their names were masked and the only available information was the number of reviews and the average review score. Participants indicated which option they preferred by pressing the respective button. They made 200 consecutive choices.}
\vspace{-4mm}
\end{figure}

In this study, we present a novel experiment intended to reveal user preferences in two widely used online websites---Amazon and IMDb. Specifically, we presented participants with options characterized by two informational social attributes: the average score and the number of reviews. We removed all other information about the options to focus on how people trade off these two pieces of information. We then tested the ability of different functions to describe user preferences and assessed preference variability and the potential for personalization of preferences in the user population. The current method can be seen as complementary to traditional learning to rank techniques that rely on logged data  \cite{joachims2002optimizing}. 

\section{Experimental design}

\begin{figure*}[htbp!]
\includegraphics[width=\linewidth, height=0.27\textheight]{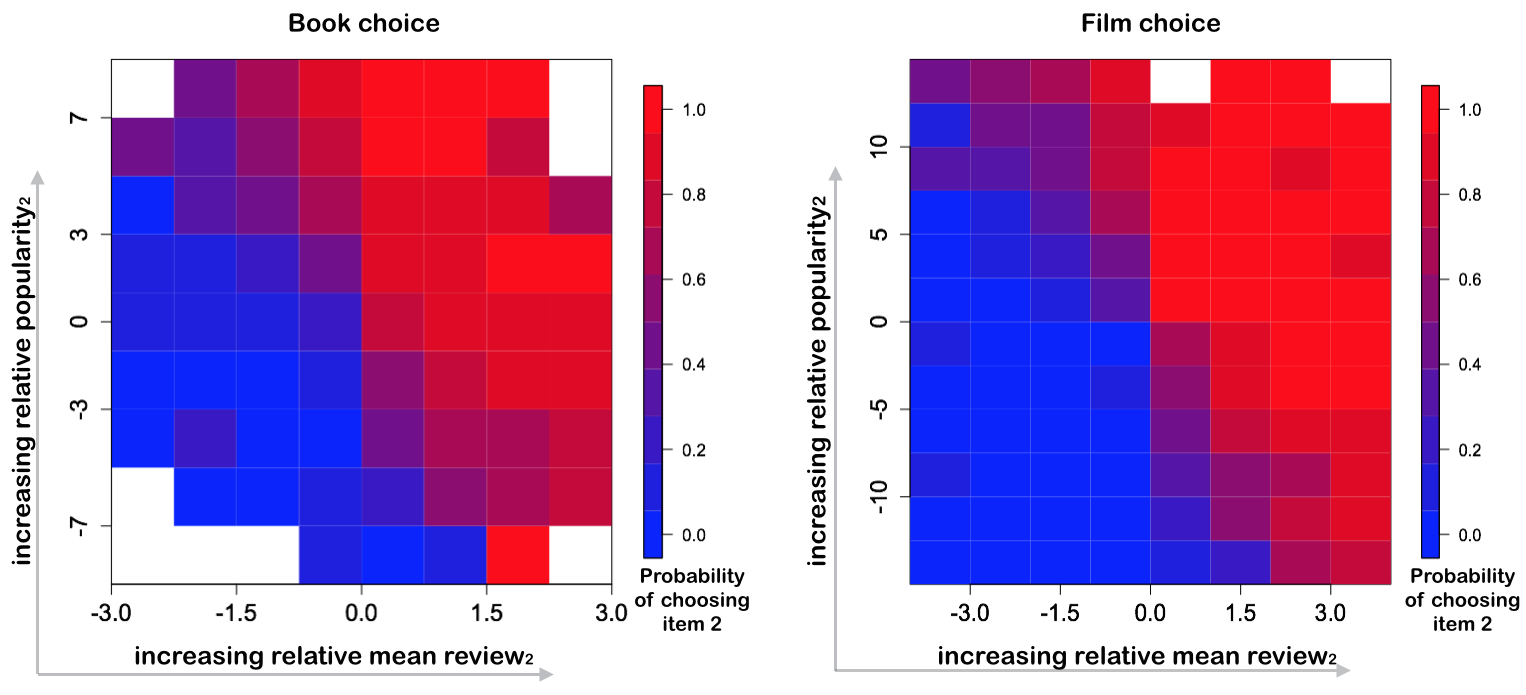}
\caption{Summary of all choices made in each experiment. The x axis shows the differences in average scores between the options. Positive numbers indicate that the option presented on the right side of the screen had a higher value. The y axis shows a logarithmic transformation (base = 2) of the ratio of the number of reviews. Positive numbers signify that the option presented on the right had more reviews. The color coding indicates the probability with which the left or right item was selected in different types of match-ups. White spots indicate an absence of data. In both experiments, people took into account popularity information in close match-ups. This effect was more accentuated for books.}
\end{figure*}

\subsection{Online venues and stimuli}

We selected the stimuli directly from two widely used online websites: Amazon for books and IMDb for movies. For the book sample, we selected 83 fiction books released between 2000 and 2011 from the website www.amazon.com.  We sampled the books on the basis of two attributes: the number of reviews and the average review score. Ratings were given on a scale from 1 to 5, with 5 being the best.  For the movie sample, we selected 98 feature movies released between 2000 and 2010 from the website www.imdb.com. We again sampled the movies on the basis of two attributes: the number of reviews and the average score.  Ratings were given on a scale from 1 to 10, with 10 being the best. We sought to create informative samples including items from the entire feature range. We used cut-off values to divide the feature space into different feature strata. This resulted to 16 different strata for books and 20 different strata for movies. We then sampled an equal number of alternatives from each of the strata. 

\subsection{The choice task}

We implemented a binary choice task in which two items, A and B, were presented simultaneously along with attributes indicating their average review score and popularity.  Figure 1 shows an example trial. The experimental task comprised 200 binary choices; each participant was presented with a different choice set randomly generated from the items in the stimuli sample. Of these 200 choices, 90\% were dilemmas and 10\% were dominated (i.e., one of the options scored higher on both attributes). The total duration of the experimental task was approximately half an hour. Overall, our design was similar to that used in risk, intertemporal choice, and inference tasks in economics and psychology.

At the beginning of the experiment, participants were given some information about the general category of products. For example, in the books experiment, they knew that the options they were to choose between were fiction books released between 2000 and 2011; they were also told that the attributes were real pieces of information retrieved from the website www.amazon.com. Similar instructions were given in the films experiment. In both experiments, the choices were consequential for participants: At the end of the experiment, participants received one item that was picked at random from the set of their choices.

\subsection{Participants and procedures}

The computerized experiments were conducted at the Center for Adaptive Behavior and Cognition of the Max Planck Institute for Human Development, Berlin, Germany. The experiments were programmed and conducted using E-Prime. The majority of the 60 participants in each experiment were German students. Participants received a show-up fee of \euro 12 plus a book or a film. In each session, participants received the same written instructions. They were seated in cubicles that inhibited visual or verbal interaction with other participants. Any questions were answered privately by one of the experimenters in the room. 


\section{Results}


\subsection{Predicting choices at the aggregate level} 


We pitted two widely implemented heuristic ranking models---relying exclusively on either popularity information or the mean review score---against models that take  possible trade-offs between these two variables into account. We used a logistic regression as our baseline model and varied its informational input to better reflect the participants' decision  processes. Finally, we compared the results with those obtained for a two-layer neural network that provided an upper performance bound for the experiments.
We used the following cross-validation process: We divided the entire dataset into a training set of 10,000 choices and a test set of 2,000 choices, disregarding the identities of individual participants. We fitted the parameters of the models in the training set and deployed the calibrated models to predict actual choices in the test set. We repeated the process 100 times and averaged the results over repetitions. The average predictive ability of the models is reported in Table 1. 

\begin{table}[htbp!]
\centering
\caption{Model performance on the aggregate data ($p$ = number of reviews; $\bar{r}$ = average score). The base for the logarithmic transformations was set to 2 throughout the analyses}
\begin{tabular}{c|c|c|c} 
 Model & Attributes & Books & Films \\ \hline 
\rule{0pt}{1.2\normalbaselineskip} Popularity & $(p_1 - p_2)$ & 0.508 & 0.58 \\ [0.8ex]
Mean review & $(\bar{r_1} - \bar{r_2})$ & 0.655 & 0.717 \\ [0.8ex]
Raw logit  & $p_1$,$p_2$,$\bar{r_1}$,$\bar{r_2}$ & 0.693 & 0.745 \\[0.8ex] 
Dim. logit & $log(p_1)$, $log(p_2)$,$\bar{r_1}$,$\bar{r_2}$  & 0.736 & 0.791 \\ [0.8ex]
Relative logit & $log(p_1/p_2)$,$(\bar{r_1} - \bar{r_2})$& 0.736 & 0.792 \\ [0.8ex]
Neural net & $log(p_1/p_2)$ , $(\bar{r_1} - \bar{r_2})$ & 0.742 & 0.791 \\ \hline
\end{tabular}
\vspace{-3mm}
\end{table}

Drawing solely on mean scores, we could predict  65.5\% of the book choices on Amazon and 71.7\% of the movie choices on IMDb. In contrast, popularity alone predicted only 50.8\% of choices in the books environment and 58\% of choices in the case of films. A logistic regression using the unprocessed average review scores and the number of reviews per option (Raw logit) improved the prediction rates for both books and and movies by more than 2.5\% as compared to merely using the average review scores. In both environments, a further improvement of more than 3.5\% over the results of the Raw logit model was obtained by using a logarithm (base = 2) of the number of reviews (Dim. logit). This finding shows that there are diminishing returns as the number of product reviews increases. 

When we further reduced the model by working with the logarithm of the ratio of the popularity scores and the difference in average review scores, the same predictive ability was maintained (Relative logit). These transformations simplify the interpretation of the models without compromising predictive power. We used the features of this model to depict all 12,000 choices made in each experiment in Figure 2. Finally, we used this informational input to train a two-layer neural network (Neural net). This model further improved prediction by small margins over the best performing models in the case of books, while it achieved similar performance in the case of movies. This indicates that the implemented variable transformations already bring the logistic regression models close to the upper bound of performance for this dataset. 

We now look directly at the logistic regression weights as a proxy of the importance of each attribute. For books, the difference in mean review scores has on average (across repetitions) 2.72 times larger weight than the logarithm of the popularity ratio (sd = 0.032). For movies, by contrast, the difference in mean review scores has on average 10.2 times larger weight than the logarithm of the popularity ratio (sd = 0.112). These results indicate that people trade off between average review scores and the popularity of items, and that the value of additional reviews is more pronounced in some market domains. 

\subsection{Learning from individual data}

People may integrate the two features at play in distinct ways. This may be due to their past experience from which they draw inferences or due to different learning and attention processes \cite{stojic2016explaining}. Alternatively, it could reflect people's inherently diverse individual preferences for popular items. The diversity of preferences at the individual level and their divergence from the behavior of the representative consumer reveal the additional benefit that can be obtained by personalizing the rankings. Ideally,  ranking algorithms should reflect the preferences of each user \cite{analytis2014multi}. To investigate the gap between aggregate and individual-level performance, we trained the models using data from every single participant in the experiment. We studied three conditions, in which we trained the models with 20, 100, and 180 choices, respectively. 

\begin{figure}[htbp!]
\includegraphics[width=\columnwidth, height=0.29\textheight]{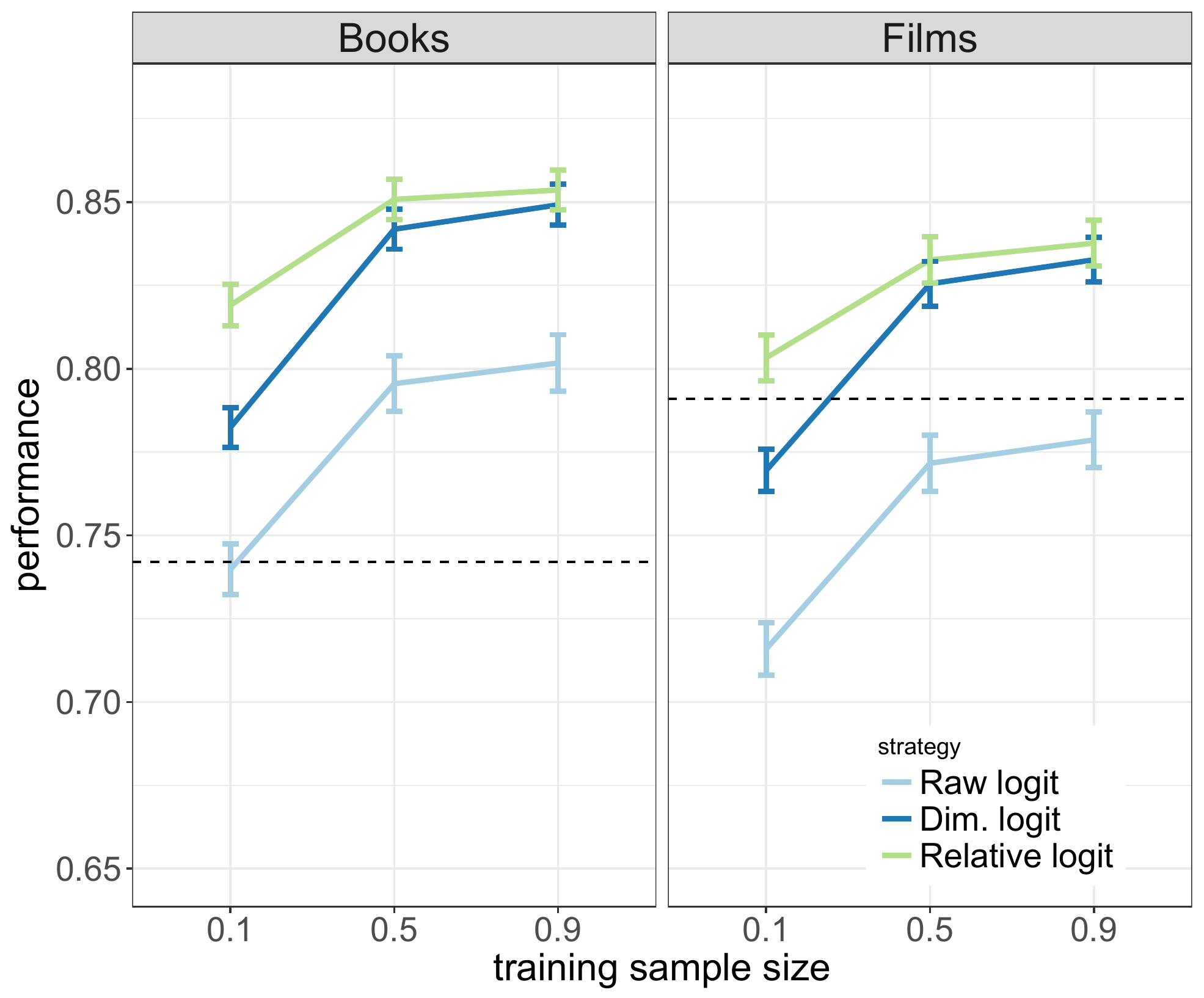}
\caption{Average performance of the various logit strategies at the individual level when the models were trained with 10\%, 50\%, and 90\% percent of the data (20, 100, and 180 choices, respectively). The dashed horizontal lines show the performance of the neural network trained on the aggregate data. With individual-level data, the models outperformed their aggregate-level counterparts by more than 10\% in the case  of books and 5\% in the case of films.}
\vspace{-4mm}
\end{figure}

As shown in Figure 3, the regression models based on a logarithmic transformation of the number of reviews already outperformed the best-performing models at the aggregate-level by large margins when sufficient data were used to train them. These findings demonstrate that there is much scope for personalizing learning in these websites. In both environments, when almost all the individual data were used, the improvement over the best-performing aggregate-level model was notable: It was more than 10\% for books and 5\% for films. Recall that the aggregate models performed better on IMDb, but this trend was reversed for individual-level data. This indicates that (i) there is larger variability of preferences in the population when choosing among books on Amazon than when choosing among films on IMDb, but also that (ii) people consistently use the same rule to make decisions. 

Indeed, we found that most people rely more on average review scores, but are willing to accept items with lower mean reviews if they are more popular (i.e. reviewed by more people). Yet, we uncovered substantial inter-individual differences. To gain a better grasp of individual-level behavior, we categorized the participants into five groups depending on the weights of the best-fitting relative logit model. As reported in Table 2, the large majority of participants in both environments had a notably larger weight on the average score than on the popularity proxy (``Rev. inclined" strategy). This difference was more pronounced in the film environment. A minority of users had a larger weight on the popularity proxy (3 in each case; ``Pop. inclined" strategy). In the books environment, one participant that had negative weights for both features (``Dissenter"). Finally, in both environments, there were small groups of participants who heuristically based their decisions on just one of the two attributes (``Popularity" or ``Mean review" strategies).

\begin{table}[htbp!]
\centering
\caption{Summary of the behavioral models describing the participants and the number of participants described by each strategy ($ w $ = weight assigned to the popularity/review attribute by the best-fitting linear model).}
\begin{tabular}{c|c|c|c} 
Strategy & Signature &  Books & Films \\ \hline 
\rule{0pt}{1.2\normalbaselineskip} Popularity & $w_p>0>w_r$ & 5 & 1 \\ [0.8ex]
Mean review & $w_r>0>w_p $ & 4 & 6 \\ [0.8ex]
Rev. inclined & $w_r>w_p>0$ & 47 & 50 \\[0.8ex] 
Pop. inclined & $w_p>w_r>0$ & 3 & 3 \\ [0.8ex]
Dissenters & $w_p<0$, $w_r<0$& 1& 0 \\ \hline 
\end{tabular}
\vspace{-2mm}
\end{table}

We identified these participants by looking for users with a negative weight for one of the two attributes. Recall that in our task 90\% of choices were dilemmas, in which participants needed to trade off between the attributes. Thus, if they consistently used only one feature, the other would appear as having a negative weight. Participants relying on a single attribute had very high levels of choice consistency, and almost all their choices could be predicted by the logit model. 
 
 \vspace{-4mm}


\section{Conclusion}

Our experiments show that people trade off between average review scores and popularity information when making choices in online interfaces. Online websites could improve their design by implementing hybrid algorithms that account for such subtle trade-offs between social features. Moreover, individual preferences vary substantially across the user population. Online venues thus stand to benefit further from learning more about how individual users integrate these two pieces of social information. 

\section{Acknowledgments}

This research was supported by NSF Award IIS-1513692 and the Max Planck Institute for Human Development.
\vspace{-2mm}
\bibliographystyle{aaai}

\end{document}